\documentclass{article}

\usepackage{cleveref}
\usepackage{amsmath}
\usepackage{graphicx}
\usepackage{comment}
\usepackage{subfloat}
\usepackage{subfig}
\usepackage{geometry}
\usepackage{tabularx}
\usepackage{listings}
\usepackage{sidenotes}
\usepackage{amssymb}
\usepackage[figuresright]{rotating}

\begin{document}

\title{Design space exploration in the microthreaded many-core architecture}

\author{Irfan Uddin\\
University of Amsterdam, The Netherlands\\
mirfanud@uva.nl}

\maketitle

\begin{abstract}

Design space exploration is commonly performed in embedded system, where the
architecture is a complicated piece of engineering. With the current trend of
many-core systems, design space exploration in general-purpose computers can no
longer be avoided. Microgrid is a complicated architecture, and therefor we
need to perform design space exploration. Generally, simulators are used for
the design space exploration of an architecture. Different simulators with
different levels of complexity, simulation time and accuracy are used.
Simulators with little complexity, low simulation time and reasonable accuracy
are desirable for the design space exploration of an architecture. These
simulators are referred as high-level simulators and are commonly used in the
design of embedded systems. However, the use of high-level simulation for
design space exploration in general-purpose computers is a relatively new area
of research.

\end{abstract}

\setcounter{tocdepth}{1}
\tableofcontents

\newpage

\section{Introduction}
\label{sn:introduction}

Simulators with high simulation speed and less complexity are desirable for
early Design Space Exploration (DSE) of the architecture. Any decision to
improve the architecture becomes more expensive and requires more effort at the
later stage and requires more effort, time and budget. DSE is performed in
all kinds of computer systems. However, in the embedded systems domain the use
of high-level simulation for DSE purposes has been accepted as an efficient
approach for more than a decade. In that sense, the DSE in embedded systems
pioneered the high-level simulation techniques. Therefore, in this paper we
will give details about DSE in embedded systems using high-level simulators. We
will explain that general-purpose computers are getting more complex and
therefore high-level simulators are also required for the DSE. Since
Microgrid is a complex architecture therefore we need to know its design space
before we present the high-level simulation techniques for the Microgrid.

The rest of the paper is organized as follows. In
Section~\ref{sn:dse_embedded_systems} we give an explanation of DSE in embedded
systems. In Section~\ref{sn:application-dep-indep-dse} we differentiate the
application dependent DSE and independent DSE. We give the DSE for the
Microgrid in~\ref{sn:dse_microgrid} and conclude the paper
in~\ref{sn:conclusion}.

\section{Design space exploration in embedded systems}
\label{sn:dse_embedded_systems}

Embedded systems perform predefined tasks and therefore have particular design
requirements. They are constrained in terms of performance, power, size of the
chip, memory etc. They generally address mass products and often run on
batteries, and therefore need to be cheap to be realized in silicon and power
efficient. Modern embedded systems, typically have a heterogeneous
MultiProcessor-System-on-Chip (MP-SoC) architecture, where a component can be
fully programmable processor for general-purpose application or a fully
dedicated hardware for time critical applications. This heterogeneity makes
embedded systems more complex and therefore designers use high-level simulator
to perform DSE at an early stage, because high-level simulators take less
effort to develop and less time in executing applications. In this section we
describe the high-level simulation technique used for the DSE in embedded
systems.

\begin{figure}

\begin{centering}

\includegraphics[width=0.6\textwidth]{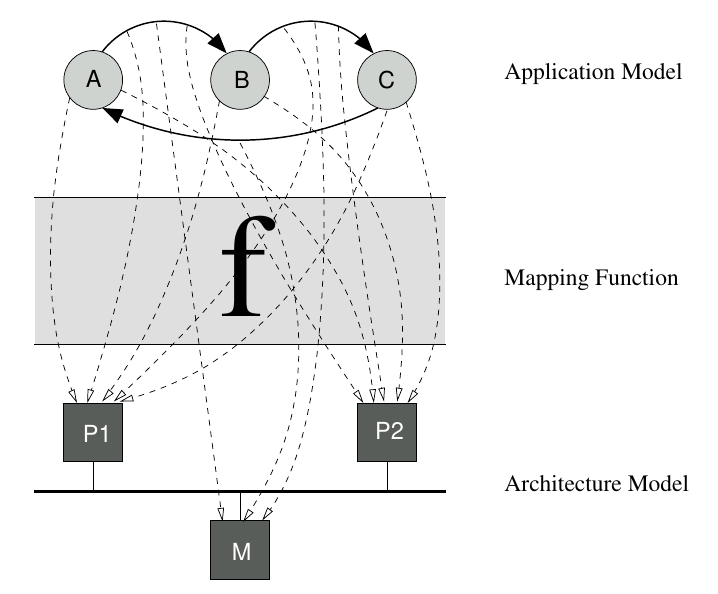}

\caption{\label{fig:mapping}The mapping of three application
processes to three architecture components resulting into a large
number of design space to be explored.}

\end{centering}

\end{figure}

Different high-level simulation techniques are introduced for the DSE in
embedded systems, and are often based on the separation of
concerns~\cite{Keutzer00system-leveldesign:} between application, architecture
and mapping functions. DSE in embedded systems is generally
application-dependent or scenario-based. Traditional embedded systems are
targeting one particular architecture and application, the aim is to explore
the design for improvement based on certain objective. Scenario-based
DSE~\cite{5647727} is the process of mapping every individual process of an
application to every architecture component with different configurations. This
mapping results in an exponential number of mapping choices i.e. design
space. We show an example in fig.~\ref{fig:mapping} (taken from
~\cite{Jaddoe:2008:SCA:1427510.1427549}) to demonstrate that only three
processes are mapped to three architecture components, but the resulted design
space is large.

Ideally DSE would like to considers all possible mappings, but an exhaustive
search is infeasible. Therefore computer architects use design pruning to
optimize the search through the design space to speed up the DSE. A smart DSE
intelligently evaluates a small fraction of the design space to come up with a
sub-optimal solution. These choices have a crucial impact on the success of the
final product. DSE addresses multiple
objectives~\cite{Erbas:2003:MOM:944645.944693} e.g. maximum performance,
minimum power consumption and less complex components. It is very difficult to
have a single solution that meets all the objectives simultaneously. The main
problem is that the objectives are conflicting e.g. low power generally means bad
performance or good performance means high power usage. Therefore a set of
solutions are selected based on a Pareto optimal front~\cite{Abido200397}, where
solutions are not dominated by any other solution looking for the same objectives.

\subsection{Related work}

High-level simulators have been used for the DSE in embedded systems domain for more
than a decade, and are used in the research of academia and industries. Below
are some of the research groups using high-level simulation for DSE in
embedded systems. There might exist other areas of research in using high-level
simulation for the DSE of embedded systems. 

\begin{itemize}

\item Sesame, University of Amsterdam~\cite{Erbas:2007:FSM:1317032.1317034}.

\item (Metro)Polis, University of California,
Berkeley~\cite{Yang_parallelsimulation}.

\item Mescal, University of California, Berkeley~\cite{mescal}.

\item Milan, University of Southern California, Los
Angeles~\cite{Bakshi01milan:a}.

\item The octopus toolset, University of
Eindhoven~\cite{Basten:2010:MDE:1939281.1939293}.

\item SystemC-based environment, STMicroelectronics~\cite{1231975}.

\end{itemize}

\section{Application -dependent and -independent DSE}
\label{sn:application-dep-indep-dse}

We want to clearly distinguish between application-dependent DSE in traditional
embedded systems and application-independent DSE in modern embedded system or
general-purpose computers. In traditional embedded systems, applications are
statically mapped to different configurations of an architecture using some
mapping functions. Based on the simulation results, innovative ideas can be
generated which can improve application, mapping and architecture separately.

In modern embedded systems we do not have one particular application or
scenario, but a range of applications targeted to a different configurations of
the architecture. For instance in smart phones it is not only one type of
application that can statically be mapped, but a range of different types of
applications are required to be explored on the different configuration of the
architecture. In a way modern embedded systems are converging to the
general-purpose systems. The range of applications increases in general-purpose
computers, where a variety of applications can be executed on the given
architecture. In these situations, the mapping of the application to
architectural component can not be analyzed statically but instead the code
patterns in algorithms are analyzed, and then different processes of an
application are dynamically mapped to different parts of the chip based on
certain objectives. Because of the dynamic mapping, application-independent DSE
is not as trivial as scenario-based DSE.

Design pruning is more structured in traditional embedded systems. For
instance, genetic algorithms, simulated annealing etc. are some of the structured
techniques that are commonly used in design pruning. However, for design
pruning in modern embedded systems or general-purpose systems, there exists
no structured solution that can dynamically determine a reduction in the design
space to optimize the search.

\section{Design space exploration in general-purpose systems}

The growing number of cores and size of the on-chip memory are creating
significant challenges for evaluating the design space of future
general-purpose computers. We need scalable and fast simulators for the
exploration of large number of cores on a chip within limited development time
and budget. Commercially available processors available in the market have few
cores on a chip e.g. Intel E708800 Series, IBM's POWER7 and AMD's Opteron 600
Series. In the near future we believe there will be hundreds of cores per chip
and DSE at the early stage can no longer be
avoided~\cite{Carlson:2011:SEL:2063384.2063454} in general-purpose computers,
as the number of mapping an application explodes as the number of cores increases.

The use of high-level simulators for the DSE in general-purpose computers is
relatively new compared to embedded systems domain. A number of simulation
techniques are in research to develop high-speed simulators for the DSE of
general-purpose computers with less complexity and shorter development time
then conventional cycle accurate simulator.
These simulation techniques are diverse and do not follow one particular
pattern. In this section we give details of some high-level simulation
techniques. There might exist other high-level simulations targeting
general-purpose computers.

\subsection{Interval simulation}

Interval simulation~\cite{Carlson:2011:SEL:2063384.2063454,GenbruggeEE10} is a
high-level simulation technique for the DSE of super-scalar single- and multi-
core processors. It raises the level of abstraction from detailed simulation by
using analytical models to derive the timing simulation of individual cores
without the detailed execution of instructions in all the stages of the
pipeline. The model is based on deriving the execution of an instruction stream
in intervals. An interval is decided based on the miss events e.g. branch
misprediction, cache misses, TLB misses etc. With interval analysis, execution
time is partitioned into discrete intervals using miss events. The analytical
models of every core cooperate with miss events in the system, and can be
extended to model the tight interleaving of threads in multi-core processors.

Interval simulation framework has two parts; functional simulation and timing
simulation, and are connected with each other through a queue. The functional
simulator feeds instructions into the tail of the queue and the timing
simulator reads those instructions from the head of the queue. The functional
simulator generates a dynamic instruction stream, including user-level and
system-level code and is subsequently fed into the timing simulator. The timing
simulator analyzes the code and advances the simulation time as per the time
required to execute an instruction stream. In case of I-cache miss, branch
misprediction and long latency load operations the simulation time is advanced
by the miss latency, branch resolution time plus the front-end pipeline depth
and long latency operations respectively.

\subsubsection*{Discussion}

Interval simulator only simulates a small number of cores in super-scalar
machines which disregards hardware microthreading and therefore the complexity
of simulating latency tolerance is not encountered. In the Microgrid we can
have more than 100 cores on the chip, and the architecture is completely
different than super-scalar machine, as it provides fine-grained latency
tolerance based on data-flow scheduling. The way programs can be written for
the Microgrid is also different. Therefore interval simulation can not directly
be used for the DSE of the Microgrid. However, we have learned some techniques
from interval simulation and have used these in HLSim. For instance, in
interval simulation in case of a cache miss the simulation time is advanced
with the addition of cache miss latency. In HLSim we advance the simulation
time with the cache miss latency but adjusted with a latency tolerance factor
based on the number of active threads. Because in case of latency tolerance the
cache miss latency can be shorter than the latency without any latency
tolerance.

\subsection{Statistical simulation}

Statistical simulation has gained interest over the past few years, as it
speeds up simulation by providing short running synthetic traces. The execution
of the original benchmarks is profiled and the key execution characteristics
are captured in a synthetic trace, which closely exhibits similar execution
characteristic as original benchmarks. The key benefit of statistical
simulation is that the synthetic trace clones the dynamic instruction count
with several orders of magnitude smaller than in the original benchmarks, and
therefore reduces the simulation time dramatically.

Nussbaum and Smith~\cite{884500} and Hughes and Li~\cite{citeulike:7771260} use
statistical simulation paradigm to evaluate multithreaded programs running on
shared-memory multiprocessor (SMP) systems. They have extended the statistical
simulation to model synchronization and accesses to shared memory. Genbrugge
and Eeckhout~\cite{Eeckhout:2003:SSA:1435585.1435761,200977} use statistical
simulation to measure some execution characteristics in the statistical profile
to be able to accurately simulate shared resources in multi-core processors.

\subsubsection*{Discussion}

Statistical simulation is a trace driven simulation technique. A synthetic
trace is generated which can be reduced to a shorter trace and is
representative of the large trace of the benchmarks. The problem with this
technique is that the original trace files can be very large which consume 
space and this technique can not consider the dynamic adaptation of multiple
applications on the chip. The high-level simulation of the Microgrid, is
execution driven i.e we dynamically generate events which are representative
of the instruction count in the basic block in a thread. These events are
mapped to the architecture and represent the execution of the application with
fine-grained interleaving. The events have information of a short piece of code
and therefore statistical simulation techniques wee not a suitable choice to be
used in HLSim.

\subsection{Sampled simulation}

The basic idea of sampled simulation is to simulate a number of sampling units
rather than the entire dynamic instruction stream. The sampling units are
selected either randomly~\cite{Conte:1996:RSL:645464.653497},
periodically~\cite{Wunderlich:2003:SAM:871656.859629} or based on phase
analysis~\cite{Sherwood:2002:ACL:635506.605403}. 

Different research in the multithreaded and multi-core processors simulation
is using sampled simulation. Van Biesbrouck et
al.~\cite{VanBiesbrouck:2004:CMG:1153925.1154587} propose the co-phase matrix
for speeding up sampled simultaneous multithreading (SMT) processor
simulation running multi-program workloads. Stenstrom et
al.~\cite{Ekman:2005:EMA:1317536.1318399} are researching the premise that fewer sampling units
are enough to estimate overall performance for larger multi-processor systems
than for smaller multi-processor system in case one is interested in
aggregate performance only. Wenisch et
al.~\cite{Wenisch:2006:SSS:1158826.1159082} have obtained similar conclusions
of throughput in server workloads. Barr et
al.~\cite{Barr:2005:AMS:1317536.1318397} proposes the Memory Timestamp Record
(MTR) to store micro-architecture state (cache and directory state) at the
beginning of the sampling unit as a checkpoint.

\subsubsection*{Discussion}

Sampled simulation is also a trace-based simulation technique which suffers from
the large trace files to be processed and changing an application results in
producing and analysing a different trace file. Every time there is some optimization
in the application, a new trace needs to be generated and analyzed.

\subsection{Related works}

There are other simulation techniques used in the design space exploration of
general-purpose computers given below.

\begin{itemize}

\item FPGA prototypes: They have low little simulation time, high accuracy and are
    useful in DSE. However these simulations require more development time and
    are more complex. They also suffer from combinatoric explosion of
    considering many low level parameters during design space exploration. Some examples
    are:~\cite{Pellauer:2008:QPM:1547560.1548281,Penry06exploitingparallelism,Chiou:2007:FST:1331699.1331723,4287395}.

\item Trace simulation: These simulation techniques generate large execution
    traces from benchmarks, and are used for the evaluation of the
    architecture. They avoid the extremely large analysis of the application,
    by executing the program only one time, generating the trace and mapping it
    to the trace to different configuration of the architecture. However a
    large storage is required in order to store the large traces and a change
    in the application requires a different trace to be generated. Statistical
    simulations and sampled simulations are some of the techniques that
    addresses the reduction of the large trace files. Some example
    are:~\cite{Conte:1996:RSL:645464.653497,IYEN96,Lafage:1999:CCT:646664.700765}.

\end{itemize}

\section{Design space exploration in the Microgrid}
\label{sn:dse_microgrid}

\subsection{Microgrid}

The
Microgrid~\cite{conf:hpc:Jesshope04,Bernard:2010:RPM:2031978.2031994,JesshopeAPC08}
is a general-purpose, many-core architecture developed at the University of
Amsterdam which implements hardware multi-threading using data flow scheduling
and a concurrency management protocol in hardware to create and synchronize
threads within and across the cores on chip. The suggested concurrent
programming model for this chip is based on fork-join constructs, where each
created thread can define further concurrency hierarchically. This model is
called the microthreading model and is also applicable to current multi-core
architectures using a library of the concurrency constructs called
\emph{svp-ptl} ~\cite{SVP-PTL2009} built on top of pthreads. In our work, we
focus on a specific implementation of the microthreaded architecture where each
core contains a single issue, in-order RISC pipeline with an ISA similar to
DEC/Alpha, and all cores are connected to an on-chip distributed memory
network~\cite{Jesshope:2009:ISM:1577129.1577136,Bousias:2009:IEM:1517865.1518255}.
Each core implements the concurrency constructs in its instruction set and is
able to support hundreds of threads and their contexts, called microthreads and
tens of families (i.e. ordered collections of identical microthreads)
simultaneously.

A number of tools and simulators are added to the designer's toolbox and used
for the evaluation of the Microgrid from different perspective. The compiler
for the Microgrid~\cite{poss.12.sl} can generate binary for different
implementations of the Microgrid. We have software libraries that provide the
run-time systems for the microthreading model on the shared memory SMP machines
and referred as \emph{svp-ptl}~\cite{SVP-PTL2009} and distributed memory for
clusters/grids and are referred as Hydra~\cite{Andrei:msc_hydra:2010} and
\emph{dsvp-ptl}~\cite{DSVP-PTL2011} The SL compiler can generate binary for
UTLEON3~\cite{5491777,danek.12},
MGSim~\cite{Bousias:2009:IEM:1517865.1518255,poss.13.MGSim.SAMOS} and
HLSim~\cite{Irfan:multipe_levels_hlsim:2013, Irfan:oneipc_hlsim:2013,
Irfan.12.2013.signatures, Irfan.12.2013.CacheBased, Irfan.01.2014.analytical,
Irfan:hl_sim_ptl:2011, Irfan:msc_hlsim:2009, Uddin:2012:CSM:2162131.2162132}.

HLSim is a high-level simulation technique aimed for the DSE of the Microgrid
and is based on discrete event simulation technique. It is execution driven
simulator and therefore does not suffer from the large size of trace files. The
events are dynamically mapped to the architecture at run time. We have built
the simulator from scratch without using any off-the-shelf code, but
some simulation techniques from Sesame and Interval simulation were used during the
development for inspiration.

\subsection{Design space in the Microgrid}

The Microgrid is a complex many-cores architecture and therefore has a huge
design space for complex application. In order to have an efficient and
validated system in the silicon we need to perform DSE in the Microgrid to
explore the performance of different applications on the different
configurations of the architecture. After DSE we can perform design pruning to
change these parameters that affect the performance. We categorize the design
space of the Microgrid as:

\begin{itemize}

\item Static architectural parameters:

\begin{enumerate}

\item Thread table size

\item Family table size

\item Frequency of cores and memory

\item Number of cores sharing an FPU

\item Frequency of delegation and distribution network

\item Size of L1-cache and L2-cache

\item Associativity of L1-cache and L2-cache

\item Number of L1-caches sharing L2-cache

\item Number of L2-caches in low-level ring

\item Number of low-level rings associated in the top-level ring

\item Distribution of address space of RAM into banks

\item Size of directory and root directory

\item The memory architecture

\item Synchronization-aware protocol

\end{enumerate}

\item Dynamic application parameters:

\begin{enumerate}

\item Place size

\item Window size

\item Cold caches

\end{enumerate}

\end{itemize}

There are some other parameters that are very low-level e.g. size of the chip,
FPU frequency, pipeline stages, the way cores are distributed on the chip etc.
We have shown only the parameters that we will simulate in the current
implementation of HLSim for the design space exploration in the Microgrid.

\section{Conclusion}
\label{sn:conclusion}

DSE is required in all kind of computer
systems. The use of high-level simulators for DSE is pioneered in embedded
systems and getting popular in general-purpose systems. As the Microgrid has a
huge design space therefore, low-level simulators are not justifiable to be
used for design space exploration. We need high-level simulators for the
efficient design space exploration. 

\section*{Acknowledgement}
The author would like to thank Dr. Raphael Poss, Dr. Michiel van Tol and Prof.
dr. Chris Jesshope.



%% References


%\bibliographystyle{elsarticle-num}
\bibliographystyle{plain}
\bibliography{main}






% \end{thebibliography}

\end{document}